\documentclass[twocolumn,showpacs,superscriptaddress,prl,aps]{revtex4}
\usepackage{graphicx}
\usepackage{epstopdf}
\usepackage{color}

\newcommand{\be}{\begin{equation}}
\newcommand{\ee}{\end{equation}}

\begin{document}
\title{Instability and evolution of the magnetic ground state in metallic perovskites GdRh$_3$C$_{1-x}$B$_x$}
\author{Abhishek Pandey}
\altaffiliation{abhishek.pandey@wits.ac.za}
\affiliation {School of Physics, University of the Witwatersrand, Johannesburg, Gauteng 2000, South Africa}
\author{A. K. Singh}
\affiliation {Department of Physics, Indian Institute of Technology, Varanasi, Uttar Pradesh 221005, India}
\author{Shovan Dan}
\affiliation {Department of Physics, University of Burdwan, Burdwan, Bengal 713104, India}
\author{K. Ghosh}
\affiliation {Department of Physics, P. C. Vigyan College, Chapra, Bihar 841301, India}
\author{I. Das}
\affiliation {Condensed Matter Physics Division, Saha Institute of Nuclear Physics, Kolkata, Bengal 700064, India}
\author{S. Tripathi}
\affiliation {Department of Physics, Indian Institute of Technology, Varanasi, Uttar Pradesh 221005, India}
\author{U. Kumar}
\affiliation {Department of Physics, National Institute of Technology, Jamshedpur, Jharkhand 831014, India}
\author{R. Ranganathan}
\affiliation {Condensed Matter Physics Division, Saha Institute of Nuclear Physics, Kolkata, Bengal 700064, India}
\author{D. C. Johnston}
\affiliation {Ames Laboratory-USDOE and Department of Physics and Astronomy, Iowa State University, Ames, Iowa 50011, USA}
\author{Chandan Mazumdar}
\affiliation {Condensed Matter Physics Division, Saha Institute of Nuclear Physics, Kolkata, Bengal 700064, India}

\date{\today}

\begin{abstract}

We report investigations of the structural, magnetic, electrical transport and thermal properties of five compositions of the metallic perovskite GdRh$_3$C$_{1-x}$B$_x$ ($0.00 \le x \le 1.00$). Our results show that all the five compositions undergo magnetic ordering at low temperatures, but the nature of the ordered state is significantly different in the carbon- and the boron-rich compositions, where the former shows signatures of an amplitude-modulated magnetic structure and the latter exhibits evidences of an equal-moment incommensurate antiferromagnetic ordering. We also observe a remarkable field-dependent evolution of conduction carrier polarization in the compositionally disordered compounds. The outcomes indicate that this system is energetically situated in proximity to a magnetic instability where small variations in the control parameter(s), such as lattice constant and/or electron density, lead to considerably different ground states. 

\end{abstract}


\maketitle

\section{Introduction}

Perovskite is one of the most well-studied crystal structure classes \cite{Pena-2001,Szuromi-2017}. Materials crystallizing in this relatively simple structure exhibit many intriguing physical phenomena as well as application-oriented properties. Usually, while referring to a perovskite compound one means $AB$O$_3$-type oxygen-containing material. However, there are several non-oxide perovskite materials that host a transition metal ion in the place of oxygen at the face-center lattice sites of the cubic unit cell, and those are usually referred to as ``{\it metallic perovskites}". MgCNi$_3$ \cite{He-2001}, RhFe$_3$N \cite{Houben-2005,Appen-2005}, Sc$_{3}M$C ($M$ = Al, Ga, In, Tl) \cite{Gesing-1997}, Mn$_3$GaC \cite{Kamishima-1998,Kamishima-2000,Tohei-2003} and $RT_{3}X$ ($R$ = rare earth ion; $T$ = Pd, Rh; $X$ = B, C) \cite{Pandey-2008a,Pandey-2008b,Takahashi-2012,Gumeniuk-2010,Sahara-2007,Music-2006,Joshi-2009,Djermouni-2011,Shishido-2001} are some of the noted metallic perovskites that have been thoroughly investigated and several exciting properties such as superconductivity, giant magnetoresistance (MR), anomalous thermal expansion, magnetocaloric effect, tunable valence behavior and temperature-independent electrical resistivity have been observed. Interestingly, several binary $RT_{3}$ compounds related to the $RT_{3}X$ perovskites also crystallize in the structurally related cubic AuCu$_3$ phase, where the body-center site remains vacant \cite{Pandey-2009a,Pandey-2009b,Pandey-2013,Elsenhans-1991, Gardner-1972}. One such compound GdPd$_3$ has recently been reported for its exciting oscillating MR behavior that arises because of the extremely fragile magnetic structure of the material \cite{Pandey-2017}. Binary GdRh$_3$, however, does not form in a melt-separable equilibrium phase \cite{Loebich-1976}, but the related GdRh$_3$B and GdRh$_3$C phases are reported to crystallize in an undistorted perovskite structure \cite{Shishido-2001,Joshi-2009}. Preliminary results reported on GdRh$_{3}X$ suggest considerably difference properties of the boron- and the carbon-containing compounds \cite{Joshi-2009}. 

We present in this paper an investigation of five compositions of GdRh$_{3}$C$_{1-x}$B$_x$ ($x$ = 0, 0.25, 0.50, 0.75, 1.00) and find a systematic evolution of the properties that is apparently linked to the changes in the lattice parameter as well as in the electron count of the system, both of which are altered when C is substituted by B. Our results indicate an evolution of the magnetic structure from amplitude modulated (AM) one in GdRh$_{3}$C to an equal-moment (EM) structure in GdRh$_{3}$B. This outcome, which is triggered by tweaking the relative compositions of small nonmagnetic metalloids B and C in a compound that hosts heavy rare-earth and transition metal elements, hints that the ground states of these phases are formed near a magnetic tipping point where a small perturbation to the parameter(s) can destabilize the energy balance and lead to a considerably different new ground state.  

\begin{figure*}[t!]
\centering
\includegraphics[width=4.5in]{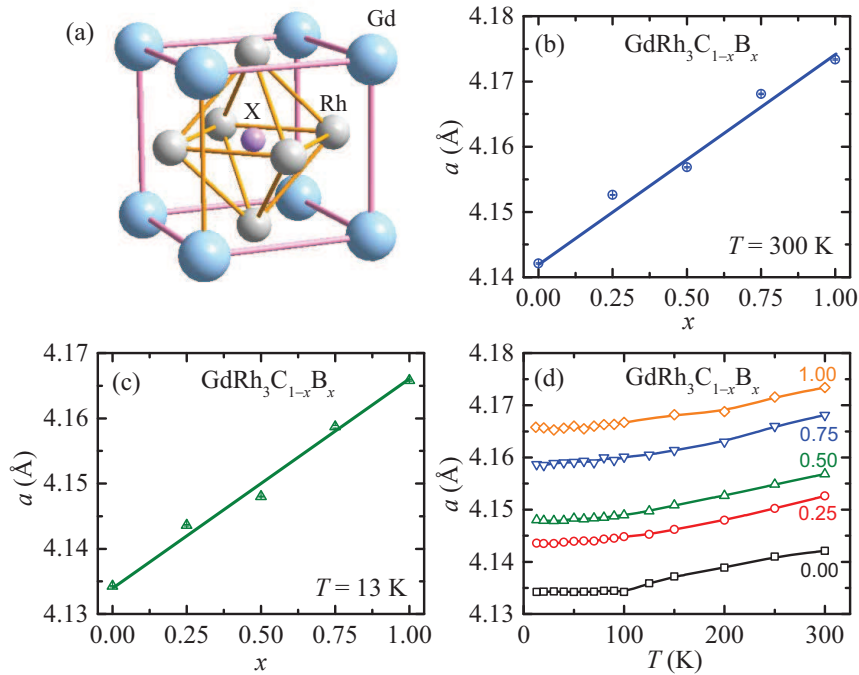}
\caption{(a) Arrangement of Gd, Rh and $X$ atoms in the cubic unit cell of GdRh$_3$$X$ ($X$ = B and C). Variation of the cubic unit cell parameter $a$ with the boron concentration $x$ at temperatures $T = 300$ and 13~K is shown in (b) and (c), respectively. Solid lines in (b) and (c) are guides to the eye. (d) The temperature dependence of $a$ of the five GdRh$_3$C$_{1-x}$B$_x$ compositions, where the $x$ is indicated next the respective plots. Sizes of the error bars in (b), (c) and (d) are smaller than the size of the symbols.}
\label{fig:LP}
\end{figure*}

\section{Experimental Details}

Samples were synthesized by arc-melting the stoichiometric amounts of highly pure (purity $\ge 99.9\%$) constituent elements from Alfa-Aesar in an inert atmosphere of argon. Post melting, the samples were wrapped in Ta foils and annealed at $1000~^{\circ}$C for 240~h  under vacuum. Structural characterization of the samples was performed by refining temperature dependent x-ray diffraction (XRD) data collected between 13 and 300~K using a TTRAX-III high-resolution powder diﬀractometer of Rigaku Inc., Japan, which is equipped with a rotating-anode Cu x-ray source. The FullProf package \cite{Carvajal-1993} was used for Rietveld refinements of the powder XRD data collected at fifteen different temperatures on the five investigated compositions. Temperature- and magnetic field-dependent magnetization $M$ measurements under zero field-cooled conditions were carried out utilizing a Magnetic Properties Measurement System of Quantum Design, Inc., USA\@. Heat capacity $C_{\rm p}$ data were collected using a Physical Properties Measurement System of Quantum Design, Inc., USA\@. Electrical transport measurements were performed between the temperatures 2 and 300~K using a home-built multisample probe installed in a cryostat supplied by Oxford Inc., UK\@.

\section{Results}

\subsection{Crystal structure and lattice parameter}

The room temperature powder XRD data of GdRh$_3$C$_{1-x}$B$_x$ ($x = 0, 0.25, 0.50, 0.75$ and 1) show that all five investigated compositions crystallize in the cubic perovskite structure (space group: $Pm\bar{3}m$, No.: 221) shown in Fig.~\ref{fig:LP}(a). Additionally, the XRD data taken at fifteen different temperatures between 13 and 300~K show that the materials remain in a single phase at all temperatures within the aforementioned temperature range (Fig.~S1). The results show that the cubic lattice parameter $a$ deduced from Rietveld analysis of the XRD data nicely follows Vegard's law \cite{Vegard-1921}, both at 13~K as well as at 300~K, where $a$ increases linearly with boron content $x$ [Figs.~\ref{fig:LP}(b) and (c)]. This observation is a confirmation of full solubility of the carbon and the boron compounds and also provides additional evidence for the single-phase nature of all the compositions. Figure~\ref{fig:LP}(d) shows that while the lattice parameters of the four compositions GdRh$_3$C$_{1-x}$B$_x$ ($x = 0.25, 0.50, 0.75$ and 1) exhibit a positive thermal expansion between 13 and 300~K, that of GdRh$_3$C displays nearly zero thermal expansion up to $\sim 100$~K, and then shows a positive thermal expansion but with a pronounced negative curvature. As the anomaly observed in GdRh$_3$C occurs at quite high temperatures, we anticipate that it has a nonmagnetic origin and is most likely related to the energy balance between the bonding strengths and/or the elastic couplings and anharmonicity of the pair potential that leads to positive thermal expansion. A result suggesting a substantial dependance of the micro-Vickers hardness on the boron content in GdRh$_3$B$_x$ compounds was reported earlier \cite{Shishido-2006}.

\begin{figure*}
\includegraphics[width=6.9in]{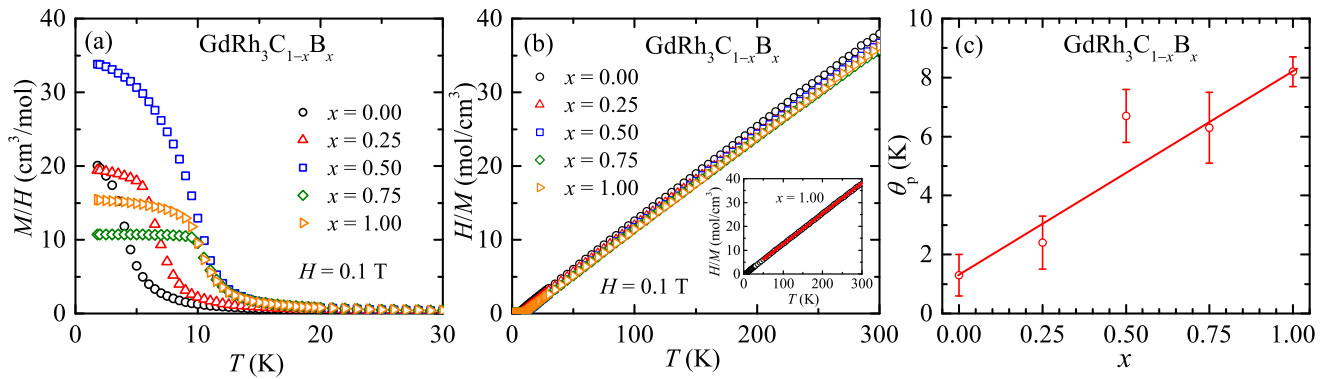}
\caption{(a) Magnetic susceptibility $\chi \equiv M/H$ versus temperature $T$ of the five GdRh$_3$C$_{1-x}$B$_x$ compositions. (b) Inverse susceptibility $\chi^{-1}$ versus $T$ plot for the same five compositions. Inset: the $\chi^{-1}(T)$ plot of GdRh$_3$C. Solid line is the Curie-Weiss fit using $\chi = C/(T-\theta_{\rm p})$, where $C$ is the Curie constant and $\theta_{\rm p}$ is the Weiss temperature. (c) Variation of $\theta_p$ with boron concentration $x$. The solid line is a guide to the eye.}
\label{fig:MT}
\end{figure*}

\begin{table*}
\caption{Parameters obtained from the analysis of powder x-ray diffraction $I(2\theta)$, magnetic susceptibility $\chi(T)$, magnetization $M(H)$, heat capacity $C_{\rm p}(T)$ and electrical resistivity $\rho(T)$ data of GdRh$_3$C$_{1-x}$B$_x$ ($x = 0.00, 0.25, 0.50, 0.75$ and 1.00) and the nonmagnetic analogue compound YRh$_3$B. The listed parameters are the lattice parameter $a$, paramagnetic Weiss temperature $\theta_{\rm p}$, effective paramagnetic moment $\mu_{\rm eff}$, magnetic moment $\mu_{\rm 5.5~T}$ at $H = 5.5$~T, highest measured value of magnetic moment $\mu_{\rm max}$ for $H \le 5.5$~T, Debye temperature $\Theta_{\rm D}$ and Einstein temperature $\Theta_{\rm E}$. The fractional contribution $u$ of the Einstein term to the $C_{\rm p}(T)$ is given in parenthesis below the respective values of the $\Theta_{\rm E}$. The magnetic ordering temperatures $T_{\rm M}$, $T_{\rm H}$ and $T_{\rm R}$ obtained from the $\chi(T)$, $C_{\rm p}(T)$ and $\rho(T)$ measurements, respectively, are listed under the column marked by $T^{*}$.}
\label{Table:Properties}
\begin{ruledtabular}
\begin{tabular}{l l l l l l l l l}
Compound & $a$ 	& $\theta_{\rm p}$ & $\mu_{\rm eff}$ & $\mu_{\rm 5.5~T}$ & $\mu_{\rm max}$ & $\Theta_{\rm D}$ & $\Theta_{\rm E}$ & $T^{*}$ \\
		 & (\AA) &	(K) & ($\mu_{\rm B}$) &	($\mu_{\rm B}/{\rm f.u.}$) & ($\mu_{\rm B}/{\rm f.u.}$) & (K)  & (K) & (K)	\\
\hline
GdRh$_3$C & 13~{\rm K}: 4.1343(1) & 1.3(7)  & 7.95(5)  & 7.01  & 7.01 & 522(21)  & 158(3) &  $T_{\rm M} = 3.3(2)$ \\
                & 300~{\rm K}: 4.1421(1)  &  &  &  &  &  & [$ u = 0.49(3)$]  & $T_{\rm H} = 3.3(1)$\\
				        &  &  &  &  &   &   &  & $T_{\rm R} = 3.3(5)$ \\
								
GdRh$_3$C$_{0.75}$B$_{0.25}$  & 13~{\rm K}: 4.1436(1) & 2.4(9) & 8.08(3) & 6.92  & 7.53 & 554(27)  & 158(3) & $T_{\rm M} = 5.5(1)$\\
                                  & 300~{\rm K}: 4.1526(1) &  &  &  &  &  & [$ u = 0.51(3)$]  & $T_{\rm H} = 5.4(1)$\\
				                          &  &  &  &  &   &  &  & $T_{\rm R} = 7.4(2)$ \\

GdRh$_3$C$_{0.50}$B$_{0.50}$  & 13~{\rm K}: 4.1480(7)  & 6.7(9) & 8.00(6)  & 6.42 & 7.08 & 526(18) & 154(3) & $T_{\rm M} = 9.0(2)$\\
                                  & 300~{\rm K}: 4.1568(2)  &  &  &  &  &  & [$ u = 0.45(3)$]  & $T_{\rm H} = 9.7(1)$\\
				                          &  &  &  &   &   &  &  & $T_{\rm R} = 10.6(1)$\\

GdRh$_3$C$_{0.25}$B$_{0.75}$  & 13~{\rm K}: 4.1587(2)  & 6.3(1) & 8.11(6)  & 6.57  & 7.23 & 517(21)  & 158(3) & $T_{\rm M} = 10.0(1)$\\
                                  & 300~{\rm K}: 4.1681(1)  &  &  &  &   &   & [$ u = 0.49(3)$]  & $T_{\rm H} = 9.9(1)$\\
				                          &  &  &  &  &   &   &  & $T_{\rm H} = 10.8(1)$\\

GdRh$_3$B & 13~{\rm K}: 4.1658(1) & 8.2(5)& 8.05(2) & 6.92 & 6.92 & 516(21)  & 157(3) &  $T_{\rm M} = 9.5(1)$\\
                & 300~{\rm K}: 4.1734(1)  &  &  &  &   &   & [$ u = 0.48(3)$] &  $T_{\rm H} = 9.4(1)$\\
				        &  &  &  &  &   &   &  & $T_{\rm R1} = 11.9(1)$\\
                &  &  &  &  &   &   &  & $T_{\rm R2} = 10.0(1)$\\

YRh$_3$B & 300~{\rm K}: 4.1647(2) & ---  & --- & --- & --- & 551(12)  & 171(2)  & ---\\
               &  &  &  &  &  &  & [$ u = 0.48(2)$]  & \\
	
	\end{tabular}
	\end{ruledtabular}
\end{table*}

\subsection{Magnetic properties}

\begin{figure*}
\includegraphics[width=6.6in]{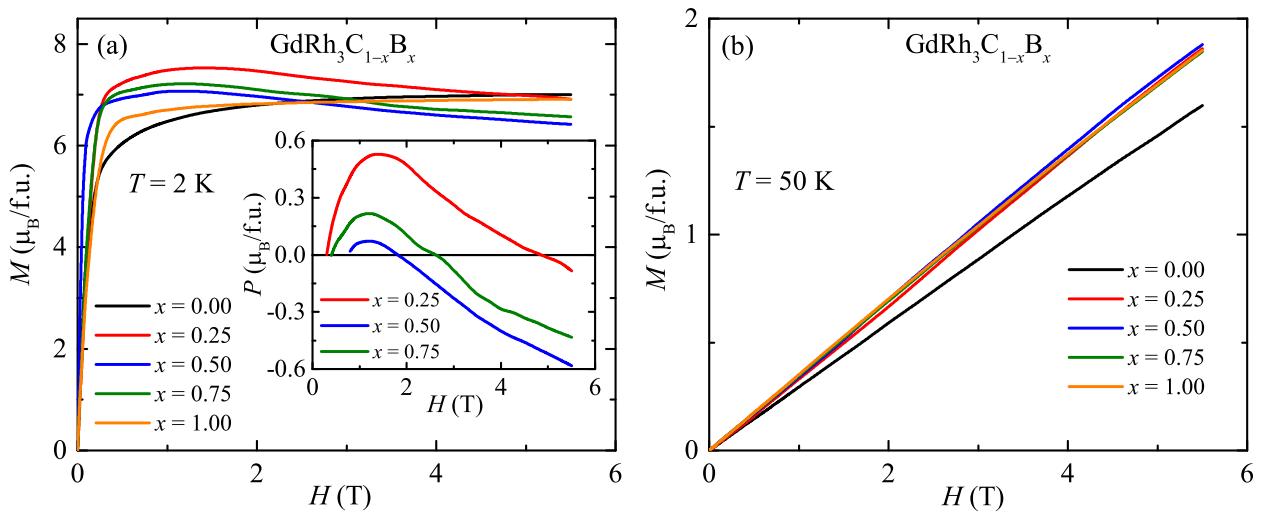}
\caption{Magnetization $M$ versus applied magnetic field $H$ plots of the five GdRh$_3$C$_{1-x}$B$_x$ compositions at $T = 2$ and 50~K are shown in (a) and (b), respectively. Inset in (a) shows the evolution of conduction electron polarization $P$ for $x = 0.25, 0.50$ and 0.75 compositions, which has been extracted by subtracting $7~\mu_{\rm B}$ from the measured $M$, with $H$.}
\label{fig:MH}
\end{figure*}

The magnetic susceptibilities $\chi \equiv M/H$ of the five GdRh$_3$C$_{1-x}$B$_x$ compounds at low temperatures ($T \leq 30$~K) are shown in Fig.~\ref{fig:MT}(a). The shapes and magnitudes of the $\chi(T)$ plots suggest ferromagnetic (FM) ordering in all five samples, but with considerably different magnetic ordering temperatures $T_{\rm M}$ that range between $\sim 3$ and 10~K (Table~\ref{Table:Properties}). The linear variation of the inverse susceptibility $\chi^{-1}$ with temperature for $T \gtrsim T_{\rm M}$ indicates the presence of local magnetic moments in the materials that follow the Curie-Wiess law in the paramagnetic $T$-region [Fig.~\ref{fig:MT}(b)]. The fitted values of the effective paramagnetic moment $\mu_{\rm eff}$ and the Weiss temperature $\theta_{\rm p}$ in the Curie-Weiss law are listed in Table~\ref{Table:Properties}. The estimated value of $\mu_{\rm eff}$ of GdRh$_3$C agrees with the expectation (7.94~$\mu_{\rm B}$) for the spin $S = 7/2$ Gd$^{+3}$ ion with $g$-factor $g = 2$. However, the values of $\mu_{\rm eff}$ in the remaining four compositions are slightly larger than expected, which apparently is due to partial polarization of the conduction carriers \cite{Harmon-1974, Roelandt-1975,Sakai-1991,Sandratskii-1993,Ahuja-1994,Tosti-2003,Pandey-2009c}. The $\theta_{\rm p}$ varies almost linearly with the boron content [Fig.~\ref{fig:MT}(c)]; a fact which shows that boron incorporation and the resultant lattice expansion manifests into a significant alteration of the strength of the resultant magnetic interaction in the compound. It also suggests that likely there are competing magnetic interactions present in these materials, and a minute tweaking of the parameters, such as, the distance between the magnetic Gd$^{+3}$ ions and/or the change in conduction carrier density induced by doping holes by replacing C with B can alter their energy balance and could possibly lead to different magnetic ground states.      

Figure~\ref{fig:MH}(a) shows isothermal magnetization $M$ versus magnetic field $H$ data of the five GdRh$_3$C$_{1-x}$B$_x$ compositions measured at $T = 2$~K below their respective magnetic ordering temperatures (Table~\ref{Table:Properties}). The $M$ of the two end compositions, GdRh$_3$C and GdRh$_3$B, monotonically increases with increasing $H$ and then saturates at high fields, exhibiting a behavior which is often observed in Gd$^{+3}$ ferromagnets. While $M(H)$ data of the two compounds are qualitatively similar, they are significantly different at the low fields where the former shows a slow saturation to $\mu = 7~\mu_{\rm B}$ but the latter attains this value, which is expected from the $S = 7/2$ Gd$^{+3}$ ions, at relatively lower field of $\approx 0.5$~T [Fig.~\ref{fig:MH}(a)]. The $M(H)$ data of the other three compositions $x = 0.25, 0.50$ and 0.75 are substantially different. The $M$ in these three compositions first increases sharply with increasing $H$, then shows a broad peak, and then starts decreasing with the further increase of $H$. The maximum moment values $\mu_{\rm max}$ that these three compositions attain are significantly larger than possible solely from the Gd$^{+3}$ ions (Table~\ref{Table:Properties}), indicating that a sizable field-dependent contribution to the measured moment is coming from polarization $P$ of the conduction carriers \cite{Harmon-1974, Roelandt-1975,Sakai-1991,Sandratskii-1993,Ahuja-1994,Tosti-2003,Pandey-2009c}. The $\mu_{\rm max}$ is largest in GdRh$_3$C$_{0.75}$B$_{0.25}$, where it attains a value of 7.53~$\mu_{\rm eff}$ (Table~\ref{Table:Properties}), suggesting that the $P$ contributes up to 0.53~$\mu_{\rm eff}$ to the observed moment in this composition in the explored applied field range [Inset, Fig.~\ref{fig:MH}]. This value is in good agreement with the polarization $0.6 \pm 0.1 \mu_{\rm eff}$ reported in literature for Gd metal \cite{Harmon-1974, Roelandt-1975,Sakai-1991,Sandratskii-1993,Ahuja-1994,Tosti-2003}. The nonmonotonic behavior of the $M(H)$ data suggests that the $P$ is strongly field dependent in these three compounds. Furthermore, $P(H)$ plot shows a crossover from positive to negative values at the sample-dependent $H$s [Inset, Fig.~\ref{fig:MH}], suggesting that along with the magnitude its orientation relative to the localized moments also evolves with $H$, which is a remarkable result. Additionally, the fact that the field-dependence of $P$ occurs only in the compositions where either boron or carbon is partially substituted hints that this phenomenon is apparently linked with the lattice disorder introduced by the substitution and the resultant modification of the local interactions. As expected in the paramagnetic state of localized moments, the $M(H)$ plots are linear at $T = 50$~K [Fig.~\ref{fig:MH}(b)]. The $M(H)$ of the GdRh$_3$C$_{1-x}$B$_x$ compounds at three different temperatures are shown in the Figure~S2 of Supplementary Materials.  

\subsection{Heat capacity and magnetic entropy}

\begin{figure}
\includegraphics[width=3.3in]{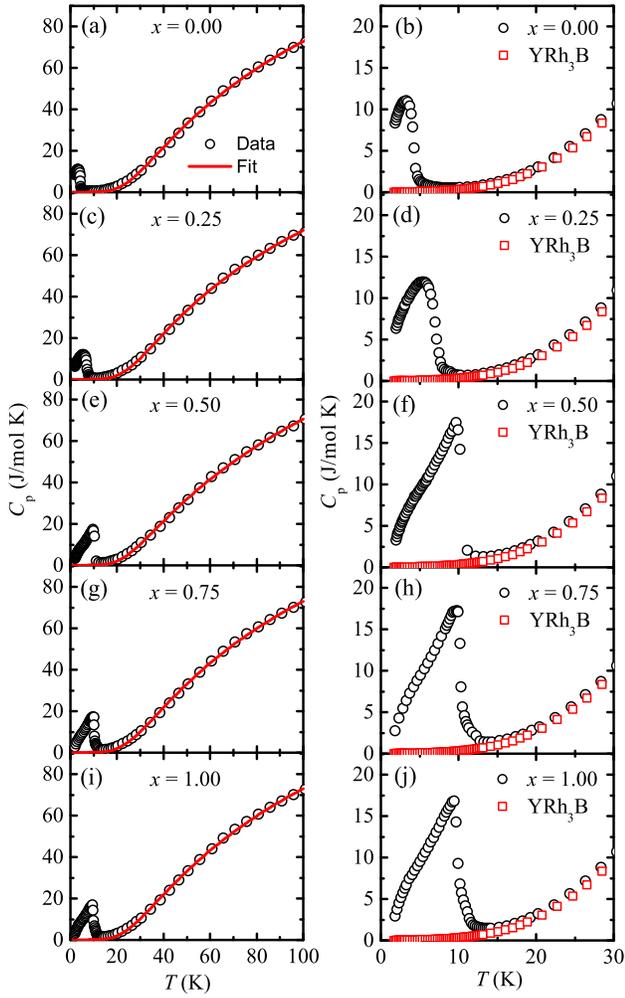}
\caption{Left panel: Heat capacity $C_{\rm p}$ versus temperature $T$ data of the five GdRh$_3$C$_{1-x}$B$_x$ compositions. Solid red curves in the left panel figures are the fits of the $C_{p}(T)$ data using the Debye-Einstein model as discussed in the text. Right panel: Low-temperature ($T \le 30$~K) $C_{\rm p} (T)$ data of the GdRh$_3$C$_{1-x}$B$_x$ compositions are shown along with the $C_{p} (T)$ data of the nonmagnetic analogue compound YRh$_3$B\@. The temperature axis of the $C_{p} (T)$ data of YRh$_3$B was scaled to incorporate the difference in its molar mass compared to the GdRh$_3$C$_{1-x}$B$_x$ compounds \cite{Pandey-2017}.}
\label{fig:HC-DE}
\end{figure}

\begin{figure}
\includegraphics[width=3.3in]{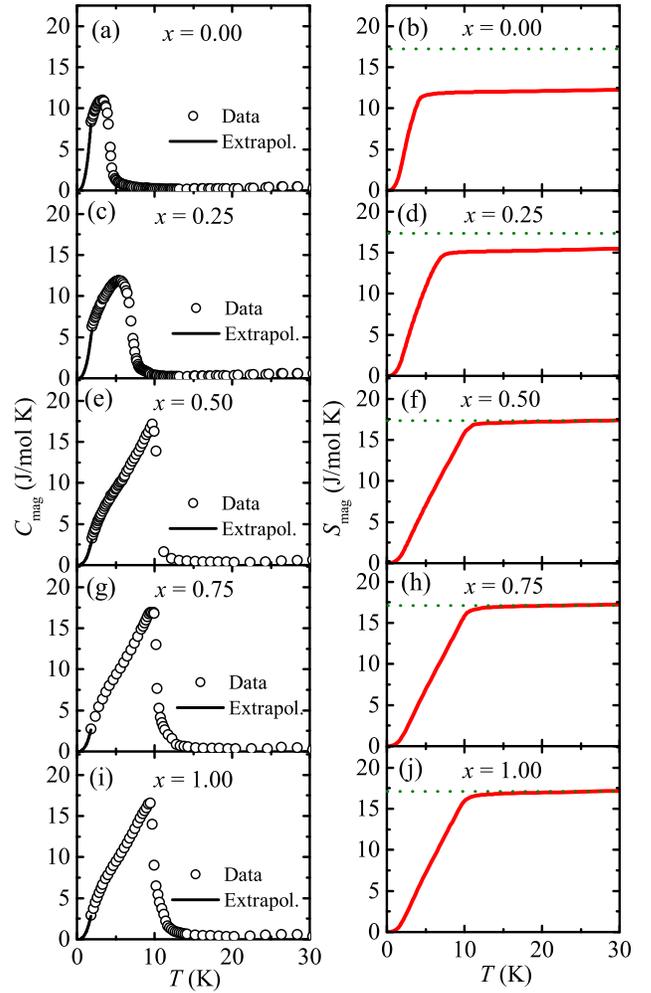}
\caption{Left panel: Magnetic contribution $C_{\rm mag} = C_{\rm p} - C_{\rm YRh_{3}B}$ to the capacity of the five GdRh$_3$C$_{1-x}$B$_x$ compositions, where $C_{\rm YRh_{3}B}$ is the heat capacity of the non-magnetic analogue YRh$_3$C. Solid curves are the extrapolation of the experimental data down to $T = 0$~K using the expression $C_{\rm mag} = BT^{3}$. Right panel: Magnetic entropy $S_{\rm mag} = \int_{0}^{T}\frac{C_{\rm mag}}{T}dT$ of the respective compositions. Dashed green horizontal lines in the figures represent the entropy $R{\rm ln}(2S+1) = 17.3$~J/mol~K associated with the Gd$^{+3}$ spins $S = 7/2$.}
\label{fig:Entropy}
\end{figure}

The heat capacity $C_{\rm p}(T)$ data of the five GdRh$_3$C$_{1-x}$B$_x$ compositions shown in the left panel of Fig.~\ref{fig:HC-DE} demonstrate the occurrence of magnetic transitions at their respective ordering temperatures $T_{\rm H}$. The high-temperature ($T>T_{\rm H}$) $C_{\rm p}(T)$ data could not be fitted satisfactorily using solely the Debye model of acoustic phonons (Fig.~S3, Supplementary Material), but the fit improved significantly by employing the single-frequency Einstein term along with the Debye model \cite{Pandey-2018} (left panels, Fig.~\ref{fig:HC-DE}). The estimated Debye temperature $\Theta_{\rm D}$, the Einstein temperature $\Theta_{\rm E}$ and the fractional contribution of the Einstein term are all listed in Table~\ref{Table:Properties}. The low-temperature ($T \leq 30$~K) $C_{\rm p}(T)$ data of the GdRh$_3$C$_{1-x}$B$_x$ compositions are shown along with the $C_{\rm p}(T)$ data of the nonmagnetic reference compound YRh$_3$B in the respective right panels of Fig.~\ref{fig:HC-DE}. The data for the boron-rich compositions ($x = 0.50, 0.75$ and 1.00) show a narrow peak centered at $T_{\rm H}$ and a shoulder at lower temperature. This behavior is expected from an equal moment (EM) local moment system under the mean-field model \cite{Bouvier-1991,Blanco-1991,Goetsch-2013,Johnston-2015}. However, the  $C_{\rm p}(T)$ data of the carbon-rich compositions ($x = 0.00$ and 0.25) exhibit a significantly different behavior in that they show a broad peak centered at their $T_{\rm H}$, which is qualitatively different from the $\lambda$-shaped peaks observed in the three boron-rich compositions. Such broad transitions have been reported in several Gd-based materials and are a characteristic signature of amplitude modulated (AM) magnetic structures \cite{Bouvier-1991,Blanco-1991,Malachias-2018,Opagiste-2019}. Molecular field theory predicts that the discontinuity in $C_{\rm mag}$ for equal moment systems with a spin $S = 7/2$ at the magnetic ordering temperature is $\Delta C_{\rm mag} = 21.14$~J/mol~K \cite{Blanco-1991,Sangeetha-2018}. The observed $\Delta C_{\rm mag}$ is 16.6(3) and 11.5(5)~J/mol~K in the boron- and the carbon-rich compositions, respectively. The smaller than expected $\Delta C_{\rm mag}$ observed in the boron-rich compositions is likely due to the presence of substantial short-ranged correlations that start building up above the respectivee $T_{\rm H}$s [Fig.~\ref{fig:HC-DE}].  

The magnetic contributions to the heat capacities estimated using $C_{\rm mag}(T) = C_{\rm p}(T) - C_{\rm YRh_3B}(T)$ are shown in the left panels of Fig.~\ref{fig:Entropy}. The $C_{\rm mag}$ below 1.8~K was extrapolated using $C_{\rm mag} = BT^3$, applicable for antiferromagnetic (AFM) spin waves \cite{Gopal-1966}, where $B$ is a constant. The magnetic entropy $S_{\rm mag}$ of the boron rich compositions ($x = 0.50, 0.75$ and 1.00) saturates to $R{\rm ln}(2S+1) = R{\rm ln}8 = 17.3$~J/mol~K at around 10~K as expected for a $S = 7/2$ system (Fig.~\ref{fig:Entropy}), where $R$ is the molar gas constant. However, the $S_{\rm mag}$ of the carbon rich compositions ($x = 0.00$ and 0.25) appear to saturate at values which are significantly smaller than $R{\rm ln}8$. This observation indicates that not all the Gd spins are ordering at the respective $T_{\rm H}$s of these compositions, and there is likely a residual entropy confined at the lower temperatures. Thus, our $C_{\rm mag}(T)$ and $S_{\rm mag}(T)$ data collectively indicate that underlying magnetic ground state in GdRh$_3$C$_{1-x}$B$_x$ compounds undergoes a notable transformation between $x = 0.25$ and 0.50.  

\begin{figure}
\includegraphics[width=3.3in]{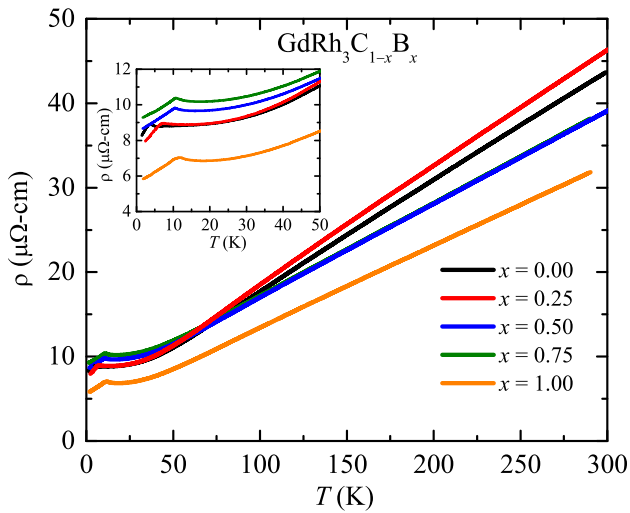}
\caption{Electrical resistivity $\rho$ versus temperature $T$ of the five GdRh$_3$C$_{1-x}$B$_x$ compositions. Inset: The $\rho(T)$ data at low temperatures $T\le 50$~K\@.}
\label{fig:Res-All}
\end{figure}

\begin{figure}[t]
\includegraphics[width=3.3in]{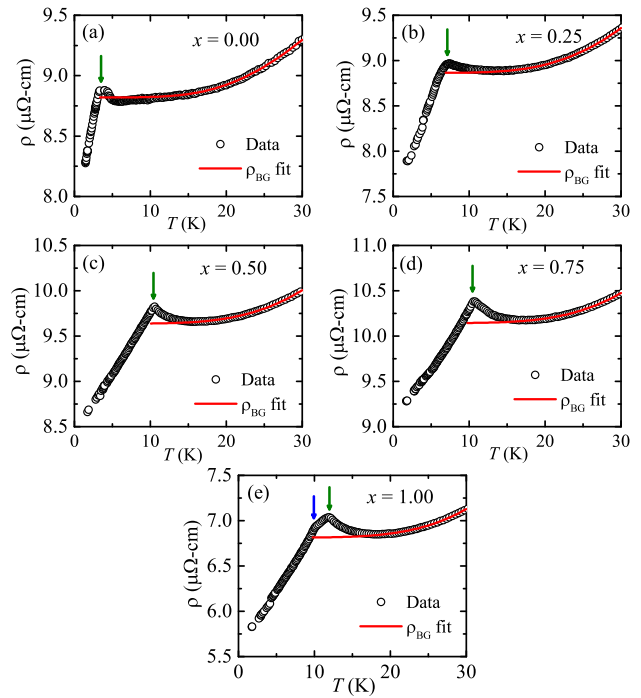}
\caption{Electrical resistivity $\rho$ versus temperature $T$ of the GdRh$_3$C$_{1-x}$B$_x$ compositions with $x = 0.00, 0.25, 0.50, 0.75$ and 1.00 at low temperatures $T\le 30$~K are plotted in (a), (b), (c), (d) and (e), respectively. The magnetic ordering temperature $T_{\rm R}$ is indicated by the vertical green arrows in (a), (b), (c) and (d). While in (e), the green and blue arrows represent $T_{\rm R1}$ and $T_{\rm R2}$, respectively. Red curves are the Block-Gr\"uneisen fits of the $\rho(T)$ data performed for $T \ge 15$~K, and then extrapolated down to low temperatures.}
\label{fig:Res-Mag}
\end{figure}

\subsection{Electrical resistivity}

\begin{figure}
\includegraphics[width=3.3in]{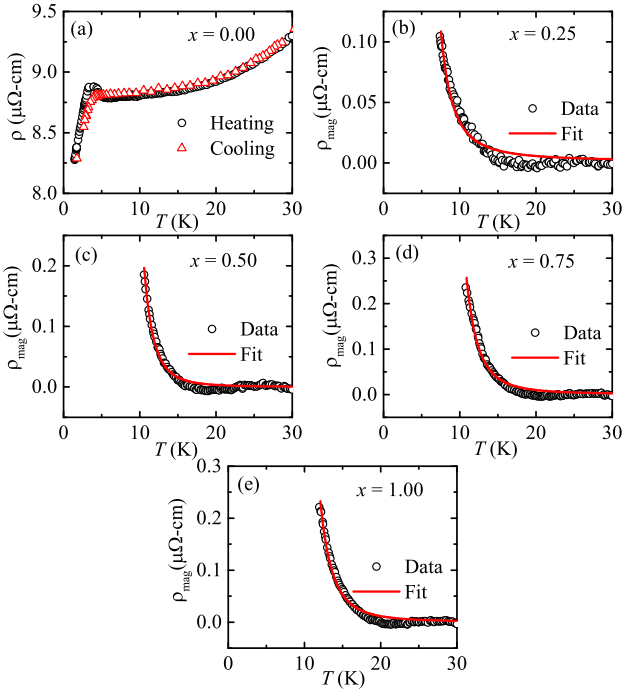}
\caption{(a) Electrical resistivity $\rho$ versus temperature $T$ of GdRh$_3$C for $T\le 30$~K for both heating and cooling cycles. The magnetic contributions to the resistivities $\rho_{\rm mag} = \rho - \rho_{\rm BG}$ versus $T$ of GdRh$_3$C$_{1-x}$B$_x$ compositions with $x = 0.25, 0.50, 0.75$ and 1.00 for $T_{\rm R} \le T \le 30$~K are plotted in (b), (c), (d) and (e), respectively, where $\rho_{\rm BG}$ is the nonmagnetic Bloch-Gr\"uneisen contribution to the resistivity as discussed in the text and $T_{\rm R}$ is the magnetic ordering temperature obtained from the resistivity measurement as indicated in by the vertical arrows in Fig.~\ref{fig:Res-Mag}. The red curves are fits using the expression discussed in the text.}
\label{fig:Res-SZG}
\end{figure}
      
The electrical resistivities $\rho$ of all five GdRh$_3$C$_{1-x}$B$_x$ compositions show metallic $T$-dependences at temperatures $T \gtrsim T_{\rm R}$ (Fig.~\ref{fig:Res-All}), where $T_{\rm R}$ is the magnetic ordering temperature deduced from the $\rho(T)$ data. The high-temperature $\rho(T)$ data were fitted satisfactorily using the Bloch-Gr\"uneisen model of electrical transport in metals (Fig.~S5). At low temperatures, the $\rho(T)$ data of the $x = 0.25, 0.50 $ and 0.75 compositions exhibit well-defined peaks at their respective $T_{\rm R}$s (Table~\ref{Table:Properties}), below which the $\rho(T)$ decreases sharply with the decrease of $T$ [Fig.~\ref{fig:Res-Mag} (b)--(d)]. The $\rho(T)$ of GdRh$_3$B shows a kink at $T_{\rm R1}$ and then an additional peak at $T_{\rm R2}$ [Fig.~\ref{fig:Res-Mag} (e)]. However, the low-$T$ $\rho(T)$ behaviors of GdRh$_3$C is different; it exhibits a small and broad hump at its $T_{\rm R}$ [Fig.~\ref{fig:Res-Mag} (a)], and then decreases sharply with the decrease of $T$. The sharp decrease in the $\rho(T)$ data observed in all five compositions below their respective $T_{\rm R}$s is due to the reduction in the spin disorder scattering in the magnetically ordered states. 

To extract the magnetic contribution $\rho_{\rm mag}$ contained in the observed $\rho(T)$ peaks of the GdRh$_3$C$_{1-x}$B$_x$ compositions, we subtracted the fitted Bloch-Gr\"uneisen resistivity $\rho_{\rm BG}$ from the experimental data (Fig.~\ref{fig:Res-Mag}). The temperature variation of $\rho_{\rm mag} = \rho - \rho_{\rm BG}$ is shown in the Figs.~\ref{fig:Res-SZG} (b)--(e) for $x = 0.25, 0.50, 0.75$ and 1.00, respectively. Because of the shallow hump observed in the $\rho(T)$ of GdRh$_3$C, it was not meaningful to do this analysis on this compound. Instead, we have plotted the thermal hysteresis data on GdRh$_3$C [Fig.~\ref{fig:Res-SZG}(a)], which shows that the shallow peak disappears when the $\rho(T)$ data are taken while cooling the sample from above its magnetic ordering temperature. This suggest the presence of some kind of spin or domain blocking accompanied with the magnetic order in this compound. A similar observation was reported earlier in magnetic measurements of GdRh$_3$C \cite{Joshi-2009}. The peaks observed in the $\rho(T)$ data of $x = 0.25, 0.50, 0.75$ and 1.00 compositions are likely due to opening of an AFM superzone pseudogap at the Fermi surface that occurs because of a magnetic structure which is incommensurate with the periodicity of the crystal structure \cite{Jensen-1991, Elliott-1964, Ellerby-1998, Park-2005, Wilding-1965, Mackintosh-1962}. To test scenario, we fitted the $\rho_{\rm mag}(T)$ data for $T_{\rm R} \le T \le 30$~K using $\rho_{\rm mag}(T) =  Ae^{\Delta/k_{\rm B}T}$, where $A$ is a constant and $2\Delta$ is the superzone band gap. We obtained reasonable fits for all four compositions. The fitted values of the parameters are listed in Table~S1.

\section{Discussion}

Our temperature dependent XRD data and their analyses confirm that all the five GdRh$_3$C$_{1-x}$B$_x$ compositions are single phase and crystallize in the cubic perovskite structure. While the $\chi(T)$ and $M(H)$ data indicate FM ordering at low temperatures, the substantial evolution of $\theta_{\rm p}$ in Fig.~\ref{fig:MT}(c) that increases roughly six times in the boron-end composition compared to the carbon-end composition, suggests that competing magnetic interactions are present in the system, and their resultant strengths vary significantly with changes in composition and lattice parameter. The slightly larger than expected value of $\mu_{\rm eff}$ and the anomalous field-dependence of $M$ observed in the $x = 0.25, 0.50$ and 0.75 compositions suggest that the conduction carriers are not only partially polarized but are also coupled in a field-dependent way with the Gd-moments. As these compositions contain both boron and carbon, this remarkable phenomenon is apparently linked with the lattice disorder and the resultant alteration of the local interactions triggered by the substitution.  

\begin{figure}
\includegraphics[width=3.3in]{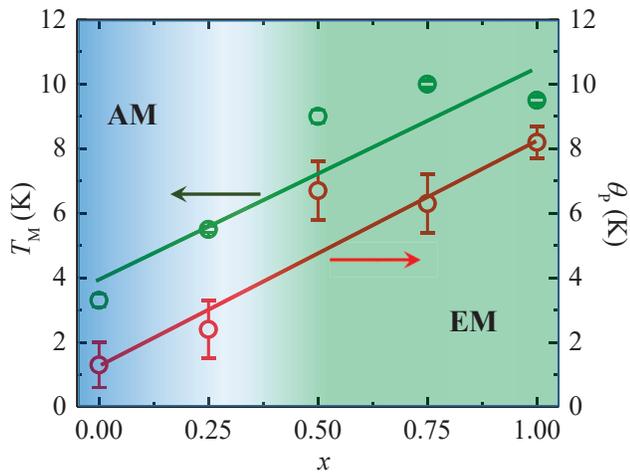}
\caption{Magnetic phase diagram of the GdRh$_3$C$_{1-x}$B$_x$ system. The left vertical axis shows the variation of magnetic ordering temperature $T_{\rm M}$ with boron content $x$ and the right vertical axis shows the variation of Weiss temperature $\theta_{\rm p}$ with $x$. The amplitude modulated (AM) and equal moment (EM) phase regions are indicated. Solid red and green lines are guides to the eye.}
\label{fig:Mag-PD}
\end{figure}

In this complex system, the $C_{\rm p}(T)$ and $\rho(T)$ data provide substantial additional information on the nature of the underlying magnetic ground state. The shapes of the low temperature $C_{\rm mag}(T)$ plots confirm that the magnetic ground states are quite different in the boron- and the carbon-rich compositions. While the former exhibits a $\lambda$-shaped peak expected from the equal moment (EM) systems, the latter shows a broad peak with a reduced $\Delta C_{\rm mag}(T)$ at the magnetic ordering temperature, a behavior which is reported for the amplitude modulated (AM) systems \cite{Bouvier-1991,Blanco-1991,Malachias-2018,Opagiste-2019}. In Ref.~\onlinecite{Blanco-1991}, Blanco {\it et al.} estimate that $\Delta C_{\rm mag}$ in AM systems is 2/3 of the jump observed in the same in EM systems. Consistent with their estimation, we observes a jump of $\Delta C_{\rm mag} = 16.6(3)$~J/mol K in the boron-rich compositions and $\Delta C_{\rm mag} = 11.5(5)$~J/mol K in the carbon-rich compositions. Therefore, the shapes of the $C_{\rm mag}(T)$ plots as well as the values of the $\Delta C_{\rm mag}$ both collectively suggest that the underlying magnetic structure undergoes a transformation from AM-type in $x = 0$ and 0.25 to EM-type in $x = 0.50, 0.75$ and 1.00 compositions (Fig.~\ref{fig:Mag-PD}). Furthermore, these results suggest that the GdRh$_3$C$_{1-x}$B$_x$ system is energetically positioned near a magnetic tipping point where a small change of the lattice parameter by $\approx 0.005$~\AA\ or the electron density by 0.25 $e^{-}$/unit cell induced by a change in composition can lead to a substantial alteration in the spin arrangements in their low temperature magnetic structures. 

The $S_{\rm mag}(T)$ of boron-rich compositions ($x = 0.50, 0.75$ and 1.00) saturates to $R$ln8 = 17.3~J/mol K, when the low temperature ($T\le 1.8$~K) $C_{\rm mag}$ is extrapolated as $BT^3$ assuming the AFM spin wave behavior [Figs.~\ref{fig:Entropy}(f), (h) and (j)]. However, if we consider a FM spin wave dispersion and extrapolate low-$T$ $C_{\rm mag}$ as $BT^{3/2}$, then $S_{\rm mag}(T)$ of the same three compositions saturates to the values which are considerably higher than $R$ln8 (Fig.~S4, Supplementary Materials). This shows that the AFM spin wave description is the correct for these three compositions.
One can employ a more accurate expression given in Ref.~\cite{Johnston-2015} for the description of low temperature $C_{\rm mag}$. However, since the simplified $BT^{3}$ provides an precise estimate at low temperatures, we have used it in our calculations. As the $\theta_{\rm p}$ decreases with increasing carbon content, it seems more appropriate to use the same AFM spin wave formalism for the extrapolation of low temperature $C_{\rm mag}$ of the carbon-rich ($x = 0$ and 0.25) compositions. However, if we do that, then the $S_{\rm mag}(T)$ saturates to the values which are significantly smaller than $R$ln8 [Figs.~\ref{fig:Entropy}(b) and (d)], suggesting that the AFM spin wave approximation does not describe the magnetic excitations down to the lowest temperatures in the carbon-rich compositions that show signatures of AM magnetic behavior. Furthermore, even the FM spin wave description does not account for the lost entropy of GdRh$_3$C (Fig.~S4, Supplementary Materials). These observations further infer that not all the Gd-spins are ordering at the respective $T_{\rm H}$s of the carbon-rich compositions. This behavior is typical to the AM systems where the amplitude of the ordered magnetic moment varies in a periodic manner right below the magnetic ordering temperature and often evolves to an EM-type structure at lower temperatures \cite{Blanco-1991}. While neutron diffraction measurements are essential in finding out the exact magnetic structure and its evolution below the magnetic ordering temperature, our $C_{\rm mag}(T)$ and $S_{\rm mag}(T)$ data together do provide indications that the underlying magnetic structures in boron-rich compositions ($x = 0.50, 0.75$ and 1.00) are EM-type while the same in carbon-rich compositions ($x = 0$ and 0.25) are AM-type (Fig.~\ref{fig:Mag-PD}).  

The GdRh$_3$C$_{1-x}$B$_x$ materials show a metallic behavior in the electrical transport measurements. The low-temperature $\rho(T)$ data of all the compositions show clear features, either in term of a broad hump observed in the case of GdRh$_3$C or well-defined peaks observed in the other four compositions. The observation of activated $T$-dependence $\rho_{\rm mag}(T) =  Ae^{\Delta/k_{\rm B}T}$ in the boron-rich compositions suggest that the underlying magnetic structure in these systems is an incommensurate AFM type (Fig.~\ref{fig:Res-SZG}). The value the prefactor $A$ goes down by a factor of about three in the carbon rich $x = 0.25$ (Table~S1). This suggests that the superzone pseudogap  starts closing with the increase of the carbon content, and the gap is fully closed with no indication of the activated behavior in GdRh$_3$C. These results confirm the conclusions achieved from the  analysis of the $C_{\rm p}(T)$ data and show that the magnetic ground states are significantly different in the boron-rich compositions ($x = 0.50, 0.75$ and 1.00) where the magnetic structure is EM-type, and in carbon-rich compositions ($x = 0.00$ and 0.25) which likely have an AM-type magnetic structure. 

Therefore, our results show that that while the magnetic measurements indicate a FM ordering in the all the five investigated compositions, the actual ground states are rather complicated and are decisively not a prototypical FM-type. Competing interaction are definitely present in the the GdRh$_3$C$_{1-x}$B$_x$ and evolve with substitution. The interplay of the interactions within the localized Gd$^{+3}$ moments as well as between the local moments and the conduction carriers lead to a variety of outcomes such as, field-dependent behavior of conduction carrier polarization, opening of superzone pseudogap and evolution of ground state from AM-modulated type to EM-type. If we consider only the two end compositions, then the magnetic ground states evolves from AM-type in GdRh$_3$C to incommensurate EM-type in GdRh$_3$B. It would be interesting to investigate if this evolution is solely triggered by the increase in the lattice parameters and hence the distance between moment-bearing Gd$^{+3}$ ions or the change in the electron density introduced by doping holes in the system by replacing C with B also has a role to play in this mechanism. In either case, this system whose magnetism originates from the $S$ state Gd$^{+3}$ ions that do not have the added complexity of the crystalline electric field effects or the Kondo/heavy fermion behaviors, presents an example of a scenario where the resultant magnetic ground state depends on a delicate balance of the parameters that are influenced by manipulating the relative content of small and non-magnetic entities.   
  
\section{Conclusion}

Metallic perovskite materials GdRh$_3$C$_{1-x}$B$_x$ show intriguing properties where perturbation caused in the lattice as well as in the electron density by manipulating the relative contents of the metalloids B and C manifests in a very significant transformation in the magnetic ground state---presumably AM-type in GdRh$_3$C to incommensurate EM-type in GdRh$_3$B. The observed noteworthy alteration in the ground state which is triggered by small nonmagnetic entities indicates that this system is positioned near a magnetic instability where small alteration in the parameters leads to considerable outcomes. Another remarkable observation is the field-dependent evolution of the conduction carrier polarization, which needs to further explored. Observation of outstanding properties in the oxygen-based as well as in the non-oxide perovskites suggest that we must keep exploring new/unexplored members of this remarkable family.\\

{\setlength{\parindent}{1in}{\bf Acknowledgments}}\\

AP acknowledges the support of Wits FRC and URC through the grants 001 254 8758101–8503 and 001.000.8758101.3113101, respectively. The work at SINP was funded by the Department of Atomic Energy, India. The work at Ames Laboratory was supported by the U.S.~Department of Energy, Office of Basic Energy Sciences, Division of Materials Sciences and Engineering. Ames Laboratory is operated for the U.S.~Department of Energy by Iowa State University under Contract No.~DE-AC02-07CH11358.

\end{document}